\pdfoutput=1
\documentclass[a4paper,british,citeautoscript,floatfix,pdftex,superscriptaddress,twoside,
aip,jcp,%
longbibliography,%
reprint%
]{revtex4-1}
\usepackage{amsfonts,amsmath,amssymb}
\usepackage{graphicx}
\usepackage[T1]{fontenc}
\usepackage[utf8]{inputenc}
\usepackage{microtype}
\usepackage[textsize=scriptsize,obeyFinal]{todonotes}
\usepackage{xspace}
\usepackage{hyperref}
\usepackage{hypernat}
\usepackage[todos]{CMImacros}

\makeatletter
\let\@fnsymbol\@fnsymbol@latex
\@booleanfalse\altaffilletter@sw
\makeatother

\newlength{\figwidth}
\newcommand{\apcn}{\ensuremath{\text{Ac-Phe-Cys-NH}_2}\xspace}%
\newcommand{\methods}{\hyperref[sec:methods]{\emph{Methods}}\xspace}%

\newcommand{\cfeldesy}{\affiliation{Center for Free-Electron Laser Science, Deutsches
      Elektronen-Synchrotron DESY, Notkestrasse 85, 22607 Hamburg, Germany}}%
\newcommand{\uhhcui}{\affiliation{The Hamburg Center for Ultrafast Imaging, Universität Hamburg,
      Luruper Chaussee 149, 22761 Hamburg, Germany}}%
\newcommand{\uhhchem}{\affiliation{Department of Chemistry, Universität Hamburg,
      Martin-Luther-King-Platz 6, 20146 Hamburg, Germany}}%
\newcommand{\uhhphys}{\affiliation{Department of Physics, Universität Hamburg, Luruper Chaussee 149,
      22761 Hamburg, Germany}}%

\begin{document}
\title{Spatially separating the conformers of the dipeptide \apcn}%
\author{Nicole\ Teschmit}\cfeldesy\uhhcui\uhhchem%
\author{Daniel A. Horke}\cfeldesy\uhhcui%
\author{Jochen Küpper}\email{jochen.kuepper@cfel.de}%
\homepage{https://www.controlled-molecule-imaging.org}%
\cfeldesy\uhhcui\uhhchem\uhhphys%
\date{\today}%
\begin{abstract}\noindent% 150 words, no references
   Atomic-resolution-imaging approaches for single molecules, such as coherent x-ray diffraction at
   free-electron lasers, require the delivery of high-density beams of identical molecules. However,
   even very cold beams of biomolecules typically have multiple conformational states populated. We
   demonstrate the production of very cold ($T_\text{rot}\sim2.3~\text{K}$) molecular beams of
   intact dipeptide molecules, which we then spatially separate into the individual populated
   conformational states. This is achieved using the combination of supersonic expansion
   laser-desorption vaporisation with electrostatic deflection in strong inhomogeneous fields. This
   represents the first demonstration of a conformer-separated and rotationally-cold molecular beam
   of a peptide, and will enable future single biomolecule x-ray diffraction measurements.
\end{abstract}
\maketitle

\noindent
Proteins are the workhorses of biological functionality in living cells and are at the heart of, for
instance, the transport of oxygen, the catalysis of biochemical reactions and interactions, or the
reproduction of cells and replication of DNA. This wide-ranging functionality is enabled by the
unique and specific 3-dimensional (3D) structures of these systems. While every protein is composed
of a sequence of the 20 amino acids encoded in RNA, the exact sequence and resulting intra-molecular
interactions lead to a specific and unique 3D structure, determining a proteins functionality.
Changes in this 3D structure, such as misfolding, can dramatically alter protein function with
potentially wide-ranging consequences, such as neurodegenerative
diseases~\cite{Soto:NatRevNeuro4:49, Hsiao:Nature338:342, Bucciantini:Nature416:507}. Especially the
strong hydrogen bonding between amino acids within the sequence has a profound effect on the
resulting protein structure~\cite{Fersht:Nature314:235, Feldblum:PNAS111:4085,
   Nagornova:Science336:320, Dian:Science296:2369}. In order to study the underlying intra-molecular
and hydrogen-bonding interaction in detail, one often turns to studying isolated small peptide
fragments in the gas-phase~\cite{Robertson:PCCP3:1, DeVries:ARPC58:585, Rijs:ACIE49:2332,
   Weinkauf:EPJD20:309, Schwing:IRPC35:569}. However, even single amino acids and dipeptides often
populate several conformational states~\cite{Suenram:JACS102:7180, Dian:Science296:2369,
   Choi:JACS128:7320, Blanco:PNAS104:20183, Yan:PCCP16:10770}, \eg, rotational structural isomers,
complicating detailed analysis and the extrapolation from small model data to entire protein
complexes. Mapping the structure-function relationship of these biomolecular machines thus requires
reproducible samples in the gas-phase in well-defined initial states~\cite{Taatjes:Science340:177,
   Chang:Science342:98, Khriachtchev:JPCA113:8143, Lin:NatComm6:7012}. More generally, species- and
conformer-pure samples of peptides in the gas-phase would open the door for novel
non-species-specific experimental techniques, such as atomic-resolution diffractive imaging with
x-rays~\cite{Neutze:Nature406:752, Seibert:Nature470:78, Filsinger:PCCP13:2076, Barty:ARPC64:415,
   Kuepper:PRL112:083002} or electrons~\cite{Hensley:PRL109:133202, Yang:PRL117:153002},
attosecond-electron-dynamics experiments~\cite{Calegari:Science346:336}, or kinetic studies of the
chemical reactivity of a single conformer~\cite{Chang:Science342:98}. Such experiments inherently do
not distinguish which conformer was probed, making it very difficult or even impossible to interpret
data collected with more than one conformer present in the interaction volume.

To investigate biomolecules in the gas-phase requires their vaporisation without fragmentation or
ionisation. Laser desorption (LD) has been demonstrated as a technique to vaporise such thermally
labile molecules~\cite{Vastola:AMS4:107, Rijs:Springer:1}, and the combination with supersonic
expansion allows for rapid cooling of the desorbed molecules~\cite{Rijs:Springer:1,
   Meijer:APB51:395, Bakker:PRL91:203003, Teschmit:JCP147:144204}. However, even in such cold
molecular beams different conformers, which differ by rotations about single bonds, can coexist. In
order to produce a pure beam containing only a single conformer, we combine LD with electrostatic
deflection~\cite{Chang:IRPC34:557}. This allows the spatial separation of molecular species based on
their distinct interaction with the applied electric field. This so-called Stark effect is dependent
on the quantum-state-specific effective dipole moment and this technique has been demonstrated to
spatially separate conformers of small aromatic molecules~\cite{Filsinger:PRL100:133003,
   Filsinger:ACIE48:6900}, and for very small molecules it can even produce single-quantum-state
samples~\cite{Nielsen:PCCP13:18971, Horke:ACIE53:11965, Trippel:PRL114:103003}. Furthermore, due to
the rotational-state-dependence of the Stark effect~\cite{Filsinger:JCP131:064309, Chang:CPC185:339,
   Chang:IRPC34:557}, deflection allows the creation of very cold ($T_\text{rot}<100$~mK) molecular
ensembles. This can significantly improve the degrees of laser alignment and mixed-field orientation
of molecules in space~\cite{Holmegaard:PRL102:023001} and thus enable ensemble-averaged
single-molecule imaging~\cite{Spence:PRL92:198102, Barty:ARPC64:415}.

\begin{figure}
   \centering
   \includegraphics[width=\figwidth]{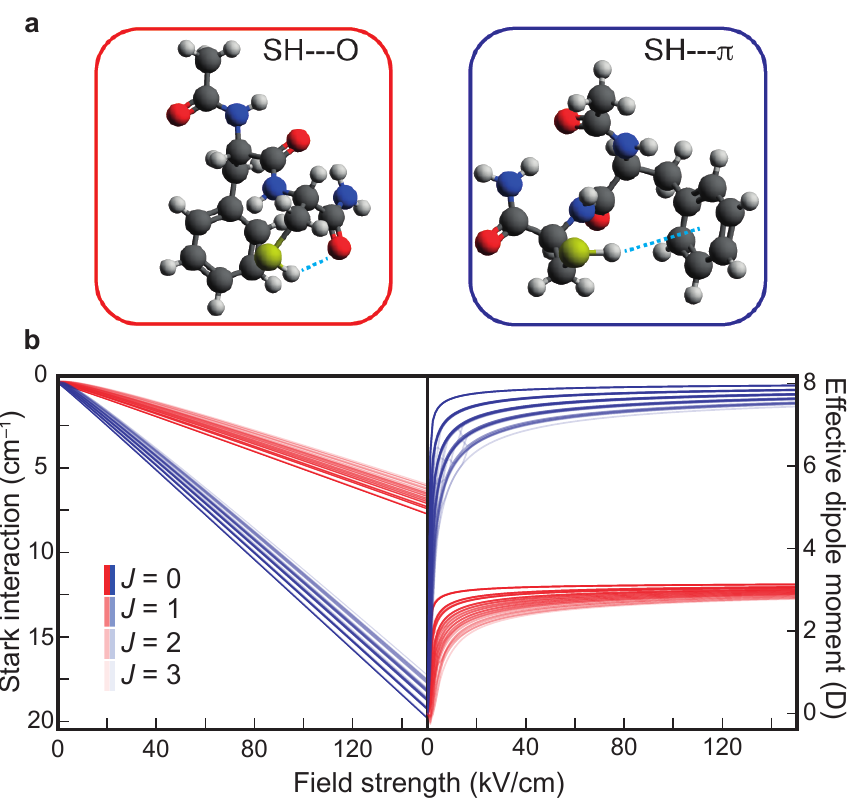}
   \caption{a: The two main conformers of the dipeptide \apcn with their distinct hydrogen-bonding
      interactions of the cystine sidechain indicated. b: The Stark energy curves (left) and
      effective dipole moments (right) for the lowest rotational states of the two conformers.}
   \label{fig:molecules}
\end{figure}
Here, we present the first combination of laser desorbed biomolecules with electrostatic deflection
and demonstrate the spatial separation of the two main conformers of the dipeptide \apcn, shown in
\autoref[a]{fig:molecules}. These two conformers differ in their hydrogen-bonding interactions and,
hence, 3D structure. One conformer, indicated by red colour throughout the paper, forms a hydrogen
bond from the SH group to the oxygen on the carboxamide group, while the other conformer, blue
colour, forms a hydrogen bond from the SH to the delocalised $\pi$-system. These two ``beautiful
molecules''~\cite{Francl:NatChem4:142} have been previously identified using vibrational and
electronic spectroscopy~\cite{Yan:PCCP16:10770}.
% need a space here otherwise the footnote command will trigger a LaTeX error
\footnote{A third conformer was detected by the Mons group~\cite{Alauddin:PCCP17:2169}, but with a
   population so minor that it is not considered in our study.} %
In a cold molecular beam these two conformers cannot interconvert, however, their significantly
different dipole moments of 3.2~D and 8.1~D result in different Stark interactions, see
\autoref[b]{fig:molecules}. This allows for their spatial separation with the electrostatic
deflector if a sufficiently cold molecular ensemble can be created~\cite{Chang:IRPC34:557}. This
would, furthermore, also separate the sample of interest from unwanted fragments or contaminants
present in the beam, such as carbon clusters from the LD process~\cite{Teschmit:JCP147:144204}.
Compared to the separation of molecular ions in ion mobility
measurements~\cite{Helden:Science267:1483, Lanucara:NatChem:6:281}, our method enables the
separation of neutral species, avoiding space-charge density limitations that severely affect
diffractive-imaging experiments~\cite{Hensley:PRL109:133202, Kuepper:PRL112:083002}. Furthermore,
the low temperatures of the generated molecular ensembles allow for strongly fixing the molecules
in space~\cite{Holmegaard:PRL102:023001} --- two prerequisites for the recording of atomically
resolved molecular movies~\cite{Filsinger:PCCP13:2076, Barty:ARPC64:415}.

\section*{Results and Discussion}
\begin{figure}
   \centering%
   \includegraphics[width=\linewidth]{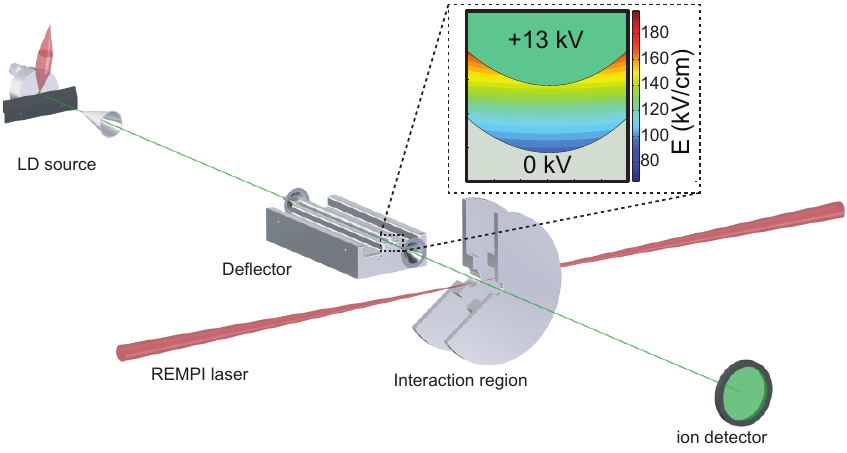}
   \caption{Schematic of the experimental setup combining the laser desorption (LD) with
      electrostatic deflection. The inset shows a cross-section of the deflector and the
      electric-field strength inside.}
   \label{fig:setup}
\end{figure}

Our implementation of the combination of LD with electrostatic deflection is shown schematically in
\autoref{fig:setup}; details are given in \methods. Briefly, the laser-desorbed molecular beam
enters a 15~cm long deflector sustaining electric field strengths on the order of
$150~\text{kV/cm}^{-1}$. The different conformers experience a different vertical deflection within
this field, which originates from the Stark-effect interaction between the molecules' space-fixed
dipole moment \mueff, \autoref[b]{fig:molecules}, and the applied electric field $\epsilon$. This
leads to a force $\vec{F}=-\mu_\text{eff}(\epsilon)\cdot\vec{\mathcal{r}}\epsilon$ acting on the
molecules~\cite{Chang:CPC185:339, Chang:IRPC34:557}. Thus, the observed deflection depends on the
effective-dipole-moment-to-mass ratio and the two conformers experience different forces, \ie,
transverse accelerations, in the electric field, leading to their spatial separation.
The molecular beam and the separation of conformers was characterised by recording spatial profiles
of the beam. This was achieved by vertically translating the ionisation laser beam through the
horizontal molecular beam, and recording the relative density as a function of laser height. The
ionisation laser was tuned to specific resonances to selectively detect a single conformer.

Such spatial molecular beam profiles for the individual conformers in the absence of an electric
field, \ie, with the deflector at 0~kV, are shown in \autoref[a]{fig:deflection_REMPI}, to which all
beam-profile intensities have been normalised. These show that both conformers are centered around
$y=0$~mm and exhibit the same spatial distribution. The measured width of the molecular beam is
predominately defined by the apertures of skimmers and the electrostatic deflector placed in the
molecular beam, see \methods. The relative population of the two conformers in the beam was assessed
by placing the ionisation laser focus at the center of the profile, as indicated by the black arrow
in \autoref[a]{fig:deflection_REMPI}, and scanning the ionisation wavelength across the
electronic-origin transitions of the two conformers around 37325~\invcm and 37450~\invcm,
respectively. The resulting resonance-enhanced multiphoton-ionisation (REMPI) spectrum is shown in
\autoref[d]{fig:deflection_REMPI} and yielded an intensity ratio of $\ordsim2:1$ for the SH--O and
SH--$\pi$ bound conformers, respectively. Assuming identical ionisation probabilities for the REMPI
process, this ratio can be taken as a measure of the relative conformer populations in the molecular
beam.

\begin{figure*}
   \centering%
   \includegraphics{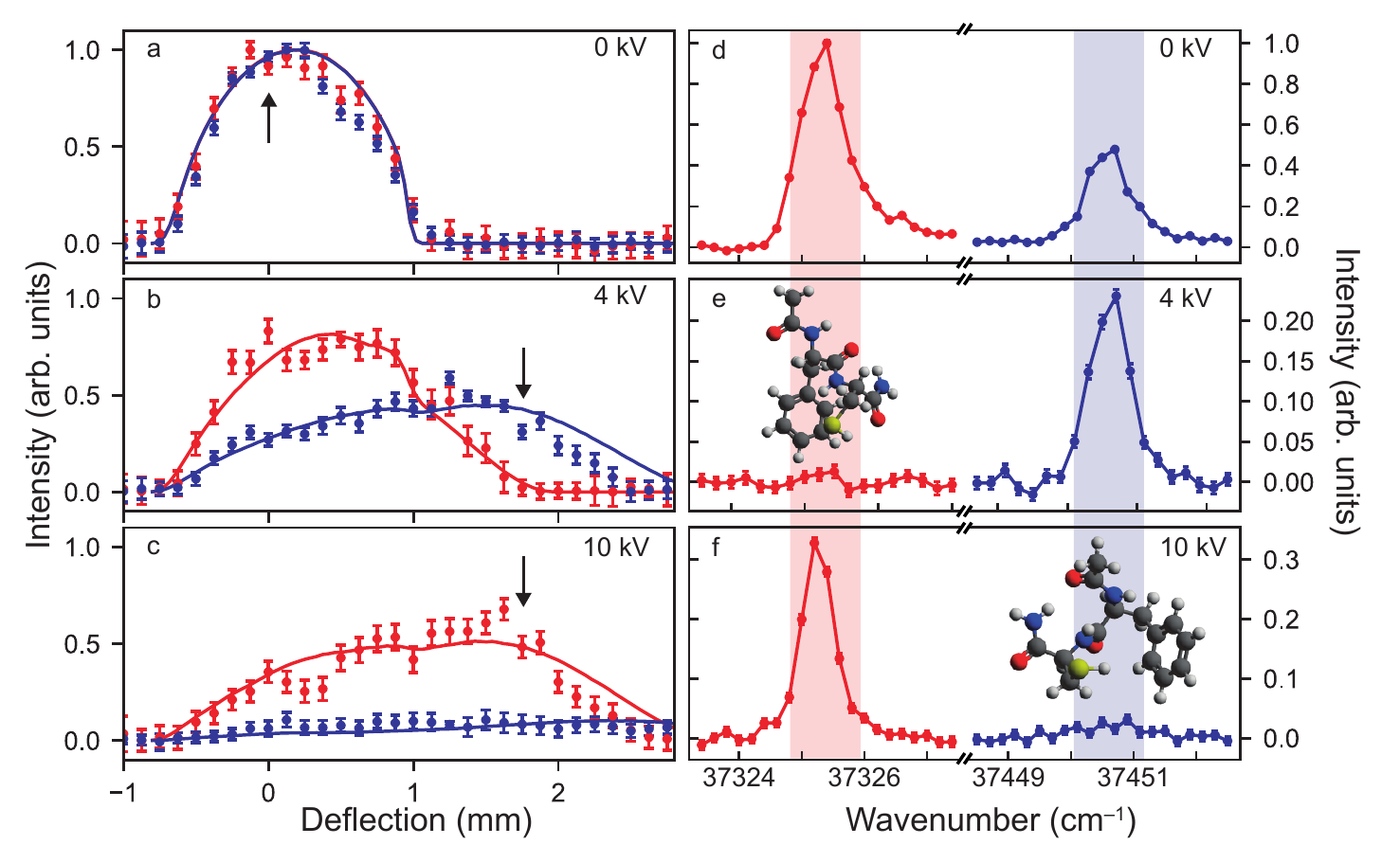}%
   \caption{Spatial molecular beam profiles (a--c) and corresponding REMPI spectra (d--f) for the
      two main conformers of \apcn. These are collected at deflector voltages of 0~kV (a,d), 4~kV
      (b,e) and 10~kV (c,e). Solid lines in the deflection profile plots (a--c) are taken from
      quantum-state resolved trajectory simulations with a 2.3~K thermal state weighting. REMPI
      spectra are taken at the spatial position indicated by the black arrow in the spatial
      profiles.}
   \label{fig:deflection_REMPI}
\end{figure*}
Charging of the electrostatic deflector lead to deflection of the molecular beam in the positive,
upward direction, as shown in \autoref[b,c]{fig:deflection_REMPI}. Application of 4~kV to the
deflector, \autoref[b]{fig:deflection_REMPI}, lead to a clear shift of both spatial profiles, with
the more polar SH--$\pi$-bound conformer shifting significantly more. This created an area, between
$\ordsim1.7$--$2.5$~mm, were a highly enriched sample of this conformer was obtained, as confirmed
by the REMPI spectrum collected at position $y=1.75$~mm and shown in
\autoref[e]{fig:deflection_REMPI}. To separate and create a pure sample of the SH--O-bound
conformer, a voltage of 10~kV was applied, leading to depletion of the SH--$\pi$-bound system from
the interaction region, as shown in \autoref[c]{fig:deflection_REMPI}. This is due to the large
deflection experienced by this more polar conformer, such that these molecules collided with the
deflector or following apertures and no clear beam was observable anymore. Instead, a
position-independent small background signal was present. A REMPI spectrum recorded in the deflected
beam is shown in \autoref[f]{fig:deflection_REMPI}, confirming the highly-enriched sample of the
SH---O-bound conformer created under these conditions.

Using a calibrated ion detector, we estimated the number of ions produced per laser shot to be
$\ordsim1$ for REMPI ionisation. By using more efficient strong-field ionisation (SFI) we extracted
a lower limit for the absolute number density of $10^7$~cm$^{-3}$, see appendix A for
details. Derivation of this density assumes an ionisation efficiency of $1$ for SFI and only takes
into account the major assigned fragmentation channels for \apcn~\cite{Teschmit:JCP147:144204} and
thus strictly represent a lower limit of the density.

Further to the deflection of the molecular beam, we observed a significant broadening of the spatial
profiles. This is due to the dispersion of the different rotational states in the electric field,
arising from the rotational-state-dependence of the Stark effect~\cite{Chang:CPC185:339,
Chang:IRPC34:557}. This is shown in \autoref[b]{fig:molecules} for $J=0\ldots3$ states, indicating
the larger effective dipole moment of lower-lying rotational states, leading to these states being
deflected more, and hence the creation of a rotationally colder sample in the deflected
beam~\cite{Filsinger:JCP131:064309, Filsinger:PCCP13:2076}. To extract approximate rotational
temperatures and quantum-state distributions in the deflected beam, we have simulated particle
trajectories through our setup for the different populated rotational
states~\cite{Chang:IRPC34:557}, details are given in appendix B. Resulting simulated
deflection profiles are shown as solid lines in \autoref[a--c]{fig:deflection_REMPI}, which were
obtained by applying a thermal-distribution weighting to the individual-state simulations,
corresponding to the rotational temperature distribution from our LD-molecular-beam source. We
extract an approximate rotational temperature of $2.3\pm0.5$~K for the laser-desorbed molecular
beam.

\begin{figure}
   \centering%
   \includegraphics[width=\linewidth]{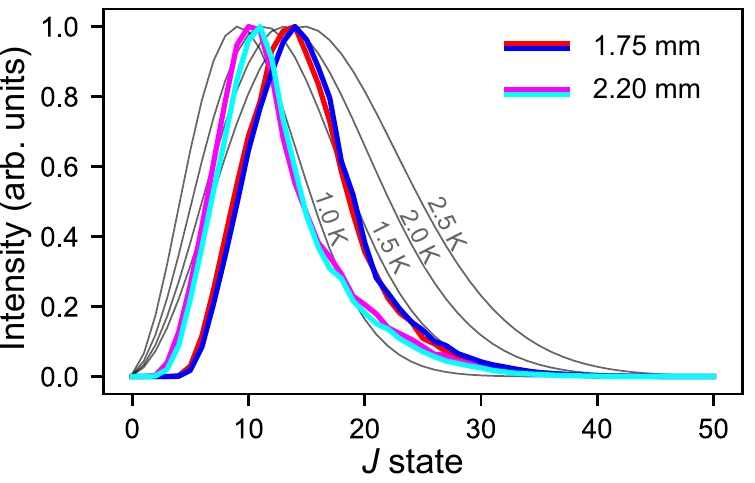}%
   \caption{Relative population of rotational states in the molecular beam for the conformer-pure
      samples at two deflected positions, 1.75~mm (as in \autoref{fig:deflection_REMPI}) and 2.2~mm.
      Shown in grey are thermal distributions at various temperatures, as indicated. The red/magenta
      and blue/cyan lines indicate distributions for the SH--O and SH--$\pi$ conformer,
      respectively.}
   \label{fig:jstates}
\end{figure}

Furthermore, we extract the quantum-state distribution within the deflected beam in
\autoref[e,f]{fig:deflection_REMPI}. These are shown in \autoref{fig:jstates} and indicate that the
deflector creates a significantly colder ensemble. While this has a non-thermal rotational state
distribution, the highest rotational states populated are approximately corresponding to a 1.5~K
distribution. Even colder ensembles can be probed by moving the interaction region further into the
deflected beam, this is indicated by the magenta and cyan distributions in \autoref{fig:jstates},
evaluated at position 2.2~mm in the deflected beam, which are comparable to a 1.0~K average.

These results highlight the quantum-state-sensitivity of the electrostatic-deflection technique,
allowing us to control conformer populations and rotational state distributions within the
interaction region and creating samples well-suited for further control techniques such as alignment
and orientation~\cite{Kumarappan:JCP125:194309, Holmegaard:PRL102:023001, Filsinger:PCCP13:2076}.
Moreover, ultrafast imaging experiments benefit from the significantly, typically several orders of
magnitude, reduced density of carrier gas in the interaction region, which does not experience
deflection in the electric field.

The presented approach is generally applicable to any polar molecule that can be vaporised by laser
desorption and entrained in a molecular beam. The achievable degree of species separation depends
crucially on the difference in dipole-moment-to-mass ratio~\cite{Filsinger:PRA82:052513,
   Chang:IRPC34:557}, and for small peptide systems we estimate that a difference of $\ordsim20$~\%
is sufficient for creating pure samples of the more polar species, whereas differences above
$\ordsim50$~\% should allow creation of a pure sample of either species, with improved setups
enabling the separation for even smaller differences~\cite{Kienitz:JCP147:024304,
   Trippel:arXiv:1802.04053}. The main limitation here is the creation of initially
rotationally-cold samples in the desorption and entrainment process, such that an appreciable
fraction of population is in the lower lying rotational states that exhibit the largest Stark shift.
If this can be further improved, for example through the use of specially designed and higher
pressure supersonic expansion valves~\cite{Even:JCP112:8068}, higher state purities or the
separation of species with smaller dipole-moment differences will be achievable.

\section*{Conclusion}
We demonstrated the combination of laser desorption for the vaporisation of labile biological
molecules with the electric deflector for the spatial separation of conformational states and the
creation of pure and rotationally-cold samples of individual conformers. Using the prototypical
(di)peptide \apcn as a model system, we showed that its two conformers, in the gas-phase, can be
spatially separated and samples of either conformer can be obtained. The measured deflection
was quantitatively understood using trajectory calculations, which furthermore allowed us to assign
a rotational temperature of $2.3\pm0.5$~K for the beam from our laser desorption source. The
generally good agreement between experiment and simulation also confirms the calculated dipole
moments and that Stark effect calculations based on the rigid-rotor approximation are sufficient
even for these large systems~\cite{Trippel:PRA86:033202}.

The created molecular samples will enable novel x-ray diffractive imaging experiments: they are
conformer-pure beams that are well-separated from carrier gas and rotationally cold enough for
strong laser alignment and orientation. The achieved densities of around $10^7$~cm$^{-3}$ are
sufficient for high-resolution diffraction experiments at free-electron laser sources such as the
European XFEL, which will deliver up to 26,000 pulses per second, allowing fast collection of data.
This enables the collection of a diffraction image within 1~h~\cite{Barty:ARPC64:415}, and simulated
aligned-molecule diffraction patterns for the two conformers, showing marked differences, are shown
in appendix C. Our laser desorption source, with its low overall repetition rate,
but reasonably long gas pulses of 100s of \us~\cite{Teschmit:JCP147:144204}, is well-suited to the
pulse-train structure of superconducting-LINAC-based XFELs~\cite{Ackermann:NatPhoton1:336}. The
produced rotationally cold samples are well suited to strong-field alignment, which can be achieved
using the available in-house laser systems available at FELs~\cite{Kierspel:JPB48:204002}.

Our developed technique will more generally enable experiments on conformer-selected biological
molecules with inherently non-species-specific experimental techniques, such as (sub-)femtosecond
dynamics~\cite{Calegari:Science346:336}, reactive collision studies~\cite{Chang:Science342:98}, or
diffractive imaging~\cite{Kuepper:PRL112:083002}. This will open new pathways to study the intrinsic
structure-function relationship of these basic molecular building blocks of the complex biochemical
machinery.

\section*{Methods}
\label{sec:methods}
% We strongly encourage authors to supply a section of up to 800 words that describes the methods
% considered to be the most essential to the paper. Additional methods, experimental procedures and
% characterization data should be placed in the supplementary information, which will be made
% available to referees during the peer- review process.
A laser desorption source, described in detail elsewhere~\cite{Teschmit:JCP147:144204}, is used to
vaporise the dipeptide \apcn (APCN, 95\% purity, antibodies-online GmbH), which is used without
further purification. The resulting cold supersonic molecular beam is skimmed twice before entering
the strong inhomogeneous field of the electrostatic deflector: once by a 2~mm skimmer (Beam Dynamics
Inc.\ Model 50.8) 75~mm downstream of the expansion, and again by a 1~mm skimmer (Beam Dynamics
Inc.\ Model 2) 409~mm downstream of the expansion. Within the strong inhomogeneous electric field of
the deflector, molecules are dispersed according to their effective dipole
moment-to-mass-ratio~\cite{Chang:IRPC34:557}. The molecular beam is skimmed once more with a 1.5~mm
skimmer (Beam Dynamics Inc.\ Model 2) prior to entering the interaction region. This skimmer can be
translated in height to ensure no part of the molecular beam is cut off. During measurements, data
is collected for two skimmer positions and subsequently combined by keeping the highest intensity
measured. The relative density of the conformers is probed via resonance-enhanced multi-photon
ionisation (REMPI)~\cite{Yan:PCCP16:10770}. The ultraviolet probe light is produced by frequency
doubling the output of a dye laser (Radiant Dyes NarrowScan, using Coumarin 153 dye in methanol),
pumped by the third harmonic of a Nd:YAG laser (Innolas, SpitLight 600). Typical laser-pulse
energies were around 19~\uJ loosely focused to a 100~\um spot in the interaction region.

The structures and dipole moments of \apcn were calculated using the GAMESS software
suite~\cite{Gordon:GAMESS:2005} using the B3LYP functional with a 6-311(p) basis set and confirmed
against published structures~\cite{Yan:PCCP16:10770}.

\vspace{1em}\noindent\textbf{Acknowledgments}\\
We thank Christof Weitenberg and the group of Klaus Sengstock for support with the wavemeter and
Thomas Kierspel for the simulation of x-ray diffraction patterns.

This work has been supported by the European Research Council under the European Union's Seventh
Framework Programme (FP7/2007-2013) through the Consolidator Grant COMOTION (ERC-614507-Küpper), by
the excellence cluster ``The Hamburg Center for Ultrafast Imaging -- Structure, Dynamics and Control
of Matter at the Atomic Scale'' of the Deutsche Forschungsgemeinschaft (CUI, DFG-EXC1074), and by
the Helmholtz Gemeinschaft through the ``Impuls- und Vernetzungsfond''. We gratefully acknowledge a
Kekulé Mobility Fellowship by the Fonds der Chemischen Industrie (FCI) for Nicole Teschmit.

\appendix
\renewcommand{\thefigure}{A\arabic{figure}}
\renewcommand{\thetable}{A\arabic{table}}
\section{Determination of a lower limit of the number density}
For the density determination a time-of-flight mass spectrum using strong-field ionisation by
femtosecond laser pulses (800~nm central wavelength, 40~fs duration, typical pulse energies of
100~\uJ) is recorded~\cite{Teschmit:JCP147:144204}. In the time-of-flight mass spectrum all peaks
originating from the \apcn molecule are integrated and the total ion current on the detector
determined. This is compared to the known calibrated current for a single ion hit, which leads to
approximately 18~ions/shot in the $\omega_0=50~\um$ focus of the laser. Assuming an ionisation
efficiency of 1 for strong-field ionisation and a molecular-beam width of 1~mm, this yields a
density of 9 x 10$^{6}$~cm$^{3}$.

\section{Numerical simulations and temperature determination}
Spatial molecular-beam profiles were simulated by first calculating the Stark energies for each
conformer for all rotational states up to $J=50$, including all states up to $J=70$ in the
calculation, using the freely available \texttt{CMIstark} software package~\cite{Chang:CPC185:339}.
Rotational constants and dipole moment vectors were taken from the DFT calculations and are
summarized in \autoref{tab:constants}.
\begin{table}[b]
   \begin{center}
      \begin{tabular}{l@{\hspace{1em}}c@{\hspace{1em}}c}
        \hline\hline
        & SH---O isomer & SH---$\pi$ isomer \\
        \hline
        A (Mhz)             & 340.181593    & 345.067516        \\
        B (MHz)             & 203.443113    & 215.965933        \\
        C (MHz)             & 159.877010    & 175.850323        \\[0.5ex]
        $\mu_\text{A}$ (D)  & 0.768         & 6.789             \\
        $\mu_\text{B}$ (D)  & 2.406         & -2.701            \\
        $\mu_\text{C}$ (D)  & 1.975         & 3.406 \\
        \hline\hline
    \end{tabular}
    \caption{Rotational constants and dipole moment vectors used for calculating the Stark effect.}
    \label{tab:constants}
  \end{center}
\end{table}
\begin{figure}
   \centering%
   \includegraphics[width=\linewidth]{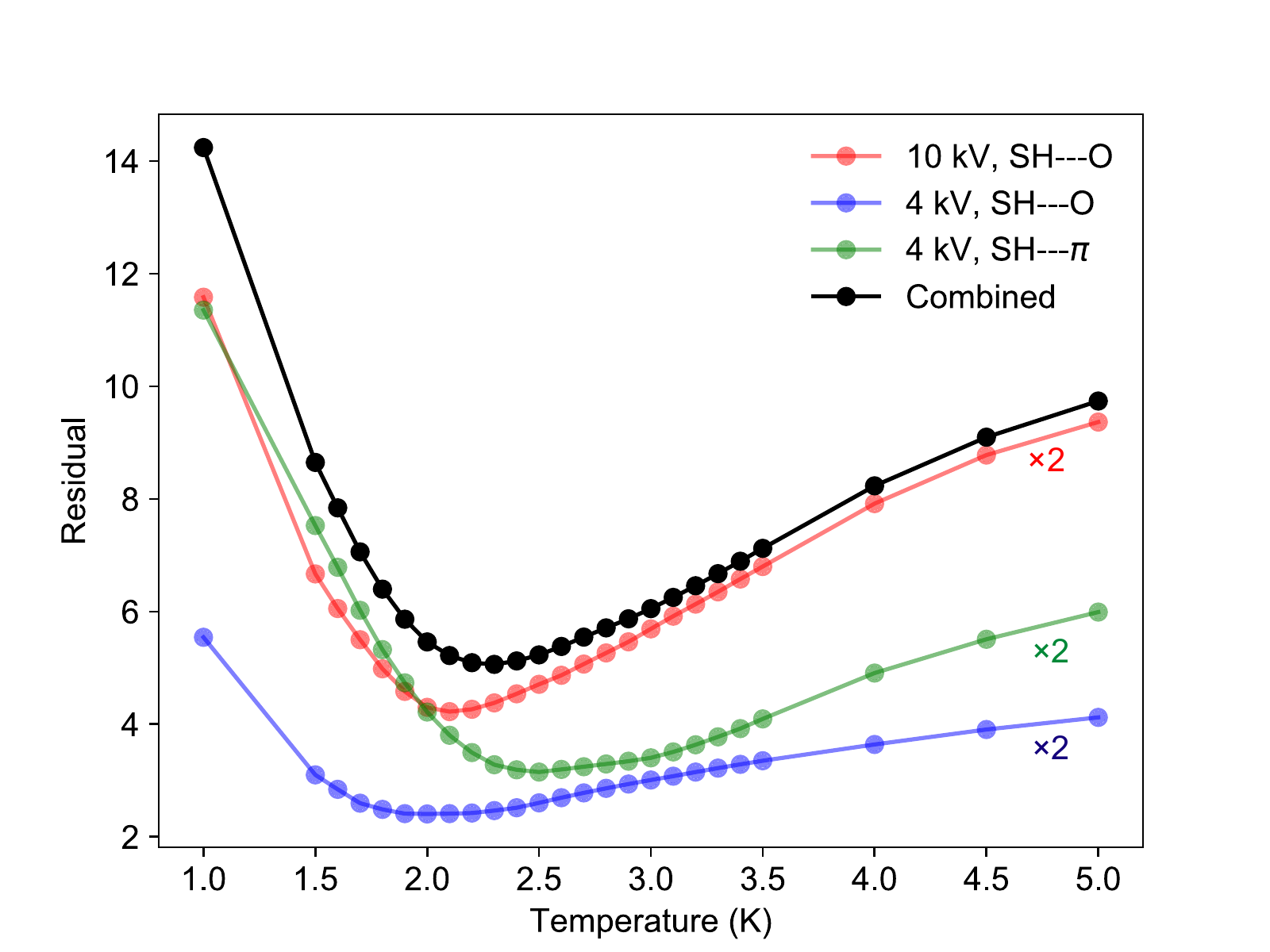}
   \caption{Residuals from fitting deflection profiles for different voltages and conformers as a
      function of temperature. Data for the SH---$\pi$ conformer at 10~kV are not shown as only a
      constant background was observed. The sum of these residuals (black line) yielded a rotational
      temperature of 2.3~K.}
   \label{fig:residuals}
\end{figure}
Subsequently, for molecules in each quantum state we carried out classical trajectory simulations
through the experimental setup, taking into account apertures and applying the appropriate forces
when molecules are within the electrostatic deflector~\cite{Chang:IRPC34:557}. Finally, histograms
of the final particle-position densities were determined at the interaction point and the
contributions from each quantum state weighted by a Maxwell-Boltzmann distribution for a given
initial temperature~\cite{Filsinger:JCP131:064309}. Simulated intensities for given conditions --
species and deflector voltage -- were scaled with a single amplitude-scaling factor to compare with
experimental data to account for additional losses and detection efficiency in the setup. The
temperature that best described the experimental observations was determined by comparing the
combined residuals, that is the absolute deviation between simulation and data, from all deflected
data sets, excluding the SH---$\pi$ conformer at 10~kV where only a constant low background was
observed experimentally, for different rotational temperatures. These are shown in
\autoref{fig:residuals} and from the combined residuals (black trace) a rotational temperature of
2.3~K for our molecular beam was extracted. Since the minima for individual deflection profiles
deviate by $\ordsim0.5$~K, conservative error bounds for the rotational temperature are $\pm0.5$~K.

\section{Simulated X-Ray diffraction patterns}
\begin{figure}
   \centering%
   \includegraphics[width=\linewidth]{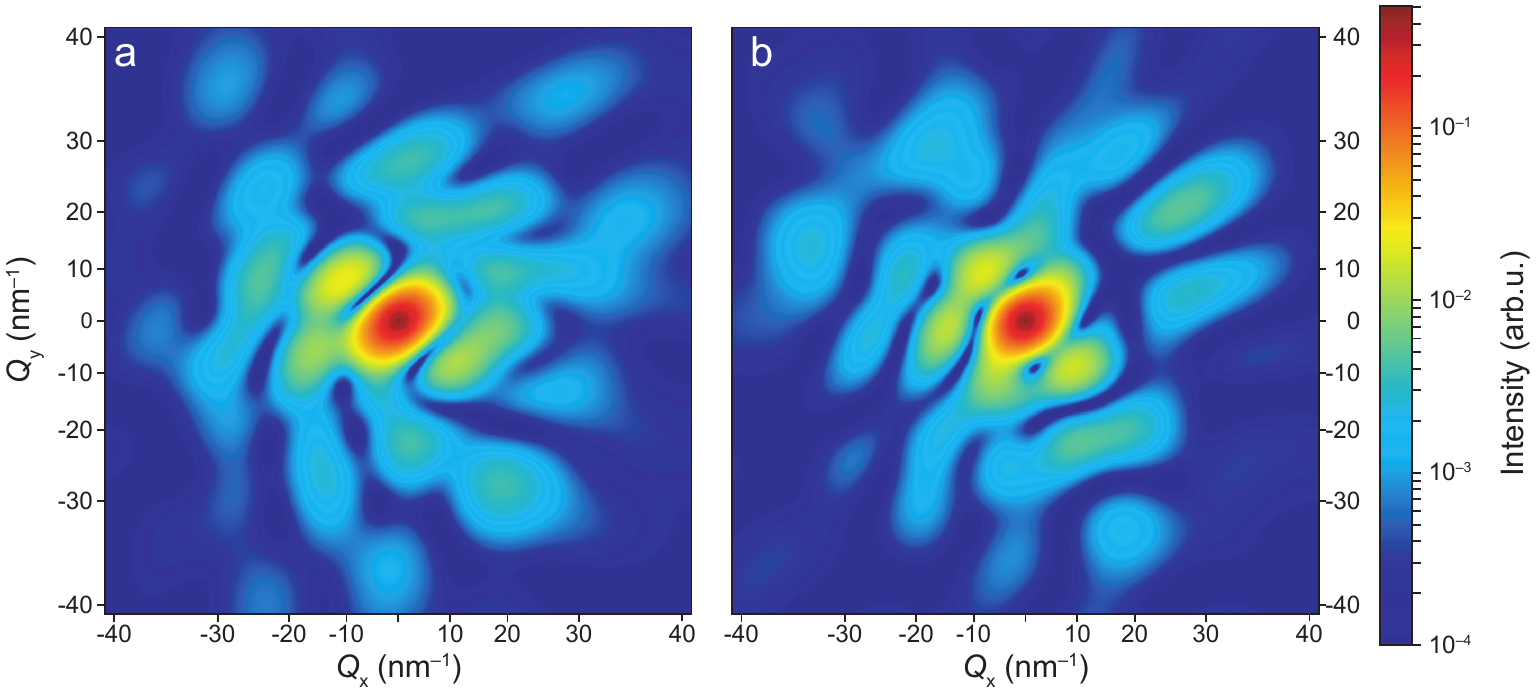}
   \caption{Simulated x-ray-diffraction patterns of the two conformers of \apcn.}
   \label{fig:Diffraction}
\end{figure}

Simulated x-ray diffraction patterns at 9.5~keV photon energy~\cite{Kuepper:PRL112:083002,
Stern:FD171:393}, achievable at current XFEL sources such as LCLS and the European XFEL, for the two
conformers of \apcn are shown in \autoref{fig:Diffraction}.
The simulation assumes a detector distance of 80~mm and maximum scattering angle on the detector of
\degree{50.2}, corresponding to a resolution of $d\approx154$~pm at the edge. These calculations
assume perfectly aligned molecules and no contribution from background gas in the interaction
region, with the most-polarizable axis of the molecules aligned vertically and the
second-most-polarizable axis aligned horizontally within the image plane, as it would be obtained in
a typical aligned-molecule-diffraction experiment~\cite{Kuepper:PRL112:083002, Stern:FD171:393,
   Kierspel:JPB48:204002}. A clear, detectable difference between the two patterns is visible.
Utilizing the 10~Hz bunch-structure of the upcoming European XFEL will enable the recording of such
patterns as well as the reaction path of conformer interconversion~\cite{Barty:ARPC64:415}.

\bibliography{string,cmi}
\end{document}